\documentclass{article}
\usepackage{spconfa4}
\usepackage{amsmath}
\usepackage{amsfonts}
\usepackage{graphicx}
\usepackage{tikz}
\usetikzlibrary{positioning}
\let\OLDthebibliography\thebibliography
\renewcommand\thebibliography[1]{
  \OLDthebibliography{#1}
  \setlength{\itemsep}{0pt}
  \setlength{\parskip}{0pt}
  \renewcommand{\baselinestretch}{0.92}
  \selectfont
}

\title{Spiking Neural Beamforming for Multichannel Speech Enhancement on the Edge}

\name{Hanan Beit-On$^{1}$ \qquad Eli Shapira$^{2}$ \qquad Boaz Rafaely$^{1}$}
\address{$^{1}$School of Electrical and Computer Engineering, Ben-Gurion University of the Negev, Israel \\
         $^{2}$Neuromorphic Lab, Israel}

\begin{document}
\ninept
\maketitle

\begin{abstract}
Speech enhancement with wearable and edge devices requires architectures that balance enhancement quality against power and memory resources. Spiking Neural Networks (SNNs) offer a promising path towards this goal through sparse, event-driven computation, but existing SNN-based speech enhancement methods are largely restricted to single-channel processing and therefore cannot exploit the spatial diversity available from microphone arrays. In this work, we propose a fully SNN framework for multichannel speech enhancement based on a fully connected architecture (FC-SNN) with sequential spike encoding, derived from a compact artificial neural network (ANN) baseline via leaky integrate-and-fire neurons. We characterize how enhancement performance scales with model size and spike-train length relative to ANN and classical beamforming baselines, and report the resulting parameter counts and computational complexity. We further compare direct and rate spike encoding, revealing a clear trade-off between event-driven computation and enhancement performance, and use these results to identify favorable configurations for edge devices.
\end{abstract}


\section{Introduction}

Speech enhancement aims to recover clean target speech from noisy acoustic environments and is a fundamental component in communication systems \cite{pandey2021dual}, hearing assistive devices \cite{han2020real}, teleconferencing \cite{Beit-On2022}, and emerging wearable platforms such as augmented reality (AR) glasses \cite{donley2021easycom, guiraud2022introduction}. While single-channel methods have achieved impressive progress with deep learning \cite{luo2019conv, hu2020dccrn, richter2023speech}, they rely mainly on spectro-temporal cues, and their performance may degrade in challenging multi-speaker scenarios.

Multichannel speech enhancement addresses this by exploiting the spatial diversity of microphone arrays, enabling significantly improved performance in reverberant and multi-speaker environments compared to single-channel approaches \cite{barker2015third, luo2019fasnet, lee2024deftan}. Recent deep neural network (DNN)-based spatial filtering and neural beamforming techniques \cite{gu2022towards, tesch2022insights, pandey2025ultra, erdogan2016improved, heymann2017beamnet, zhang2021adl, lee2024deftan1} have established the state of the art in multichannel speech enhancement. Representative approaches include mask-based beamforming \cite{erdogan2016improved, heymann2017beamnet}, all-neural spatial filtering \cite{gu2022towards, zhang2021adl}, and direct masking in the time-frequency (TF) domain \cite{lee2024deftan, lee2024deftan1}. However, these performance gains often come at the cost of increased computational and memory requirements, which can constrain their deployment on power- and resource-limited edge devices.

Spiking Neural Networks (SNNs) offer an energy-efficient alternative to conventional Artificial Neural Networks (ANNs) \cite{davies2021advancing, timcheck2023intel, yamazaki2022spiking}, through sparse, event-driven computation. Early neuromorphic approaches explored spike-based speech denoising by converting acoustic signals into spike trains and applying biologically inspired auditory processing mechanisms such as cochlear filtering and lateral inhibition \cite{wall2016,Yannan2019}. However, these methods relied primarily on handcrafted auditory processing stages and fixed network dynamics rather than data-driven training of neural network parameters, and were limited to single-channel denoising.

More recently, trainable SNN architectures for speech enhancement have been explored \cite{du2024spiking, hao2025toward, sun2024dpsnn, alohali2024temporally, riahi2023single, liu2026sse}, demonstrating the feasibility of spike-based processing for denoising tasks; however, they primarily address single-channel speech enhancement and therefore do not exploit spatial information available in multichannel recordings. In addition, some architectures combine spiking and conventional neural network components, limiting the potential energy benefits of fully event-driven computation. Furthermore, in these methods, features are typically represented using high-dimensional spike representations rather than sequential spike trains, resulting in relatively large architectures.

In this work, we address these limitations by developing a spiking neural network framework for multichannel speech enhancement that exploits spatial information while targeting efficient edge deployment. The proposed approach employs sequential spike encoding within a fully spiking architecture.

The main contributions of this work are summarized as follows:
\begin{itemize}
\item We introduce a fully spiking framework for multichannel speech enhancement that exploits spatial information from microphone-array recordings, extending SNN-based speech enhancement beyond the single-channel setting considered in previous work and providing, to our knowledge, the first exploration of multichannel speech enhancement using a fully spiking neural network.

\item We employ sequential spike encoding within a fully spiking architecture, yielding models with roughly an order-of-magnitude fewer parameters than the high-dimensional spike representations and hybrid spiking--ANN designs used in prior work.

\item Through a comparative study, we characterize how enhancement performance scales with model size and spike-train length relative to ANN baselines and traditional beamforming, identifying favorable operating points for resource-constrained deployment.
\end{itemize}

\section{Problem Formulation}

Consider a $C$-microphone array observing a target speaker in the presence of interfering speakers and background noise. The microphone signals representing the noisy observation are stacked into $C$-dimensional vectors $\mathbf{x}[n] = [x_1[n],\dots,x_C[n]]^T$, with $\mathbf{s}[n] = [s_1[n],\dots,s_C[n]]^T$ and $\mathbf{v}[n] = [v_1[n],\dots,v_C[n]]^T$, denoting the direct-path (anechoic) target speech, and the interference-plus-noise, respectively. The latter includes interfering speakers, background noise, and the target's own reverberation. The noisy observation is then given by
\begin{equation}
\mathbf{x}[n] = \mathbf{s}[n] + \mathbf{v}[n].
\end{equation}
As speech enhancement is commonly performed in the time--frequency domain, we further define the representations $\mathbf{X}[k,\ell]$, $\mathbf{S}[k,\ell], \mathbf{V}[k,\ell] \in \mathbb{C}^C$, obtained via the short-time Fourier transform (STFT) with $k$ and $\ell$ the frequency-bin and time-frame indices, satisfying
\begin{equation}
\mathbf{X}[k,\ell] = \mathbf{S}[k,\ell] + \mathbf{V}[k,\ell].
\end{equation}
The goal is to learn a mapping $\mathcal{F}: \mathbf{x}[n] \rightarrow \hat{s}[n]$ that recovers the target speech at a reference microphone. Enhancement quality is evaluated using the scale-invariant signal-to-distortion ratio (SI-SDR)~\cite{le2019sdr}, PESQ~\cite{rix2001perceptual}, and STOI~\cite{taal2011algorithm}, with SI-SDR used as the training objective.

\section{Baseline Methods}

\subsection{Baseline Beamforming Method}

An oracle minimum-variance distortionless response (MVDR) beamformer using the true steering vector and interference-plus-noise covariance matrix is used here as a baseline, with weight vector $\mathbf{w}_{\mathrm{MVDR}}[k,\ell] \in \mathbb{C}^C$ given by
\begin{equation}
\mathbf{w}_{\mathrm{MVDR}}[k,\ell]=
\frac{\mathbf{\Phi}_{vv}^{-1}[k,\ell]\mathbf{d}[k,\ell]}
{\mathbf{d}^{H}[k,\ell]\mathbf{\Phi}_{vv}^{-1}[k,\ell]\mathbf{d}[k,\ell]},
\end{equation}
where $\mathbf{\Phi}_{vv}[k,\ell] \in \mathbb{C}^{C \times C}$ is the interference-plus-noise spatial covariance matrix and $\mathbf{d}[k,\ell] \in \mathbb{C}^C$ is the target steering vector.

\subsection{Baseline Neural Network Masking Methods}

Two DNN-based masking methods are used as baselines in this work. Both methods take as input the stacked real and imaginary STFT channels across microphones ($2C$ channels), and estimate the real and imaginary parts of a complex mask $M[k,\ell]$ as two separate output channels. The first method is based on FT-JNF~\cite{tesch2022insights}, which is a recurrent ANN that applies two bidirectional LSTM layers, the first across frequency and the second across time. The estimated mask is applied to the STFT of a reference microphone as
\begin{equation}
\hat{S}[k,\ell] = M[k,\ell]X_{\mathrm{ref}}[k,\ell].
\end{equation}
The second method is a compact fully connected (FC) network (Fig.~\ref{fig:ann_snn_architectures}) that estimates the same mask by processing each TF bin independently through two ReLU hidden layers. Its input is additionally normalized per TF bin by the maximum magnitude over the channels,
\begin{equation}
\tilde{x}_c[k,\ell] = \frac{x_c[k,\ell]}{2\,\eta[k,\ell]} + 0.5, \quad \eta[k,\ell] = \max_c |x_c[k,\ell]|,
\end{equation}
mapping the input to the $[0,1]$ range required for the Bernoulli-based rate encoding of its spiking counterpart (Section~4.1). The input is further augmented with two normalized per-bin features, a frequency index and $\eta[k,\ell]$, yielding a ($2C+2$)-dimensional input shared by the FC network and its spiking counterpart (Section~4). Unlike FT-JNF, the FC architecture does not exploit temporal or cross-frequency dependencies, but its simple per-bin structure facilitates direct adaptation to a spiking implementation, as described next.

\section{Proposed Spiking Neural Network Framework}

The FC baseline is adapted to a spiking implementation, FC-SNN, by replacing ReLU activations with leaky integrate-and-fire (LIF) neurons and encoding the continuous-valued input into a spike-based representation, as illustrated in Fig.~\ref{fig:ann_snn_architectures}.

\subsection{Spike Encoding}

We consider two spike encoding schemes for converting the continuous-valued STFT features into spike-based representations. Under \textit{direct encoding}, \cite{wu2019direct,kim2022rate} each input feature $x$ (one of the $2C{+}2$ components defined in Section~3) is held constant and repeated identically at every spike timestep:
\begin{equation}
z[t] = x \quad \text{for all } t = 1,\dots,N.
\end{equation}
Because the input does not vary across timesteps, the first (input) linear layer is computed densely; only the downstream LIF layers are spike-driven. Under \textit{rate encoding}, each input value is instead converted into a stochastic spike train via a Bernoulli process with firing probability proportional to the normalized input:
\begin{equation}
z[t] \sim \mathrm{Bernoulli}(x), \quad t = 1,\dots,N,
\end{equation}
making every layer, including the first, genuinely spike-driven, at the cost of requiring substantially longer spike trains and introducing additional quantization noise. We report results under both schemes (Section~5) to characterize the resulting trade-off between the encoding scheme and enhancement performance.

\subsection{LIF Neurons}

The leaky integrate-and-fire (LIF) neuron is a computationally efficient spiking neuron model that integrates its input current over time, with a leak term causing this integrated potential to decay in the absence of input, and emits a spike once the potential crosses a threshold \cite{gerstner2014neuronal}. The input current of each LIF neuron $I[t]$ is obtained by linearly transforming the output of the preceding layer; for the first LIF layer, the output of the preceding layer is the encoded input $z[t]$ from Section~4.1. The membrane potential evolves as
\begin{equation}
u[t] = \beta u[t-1] + I[t],
\end{equation}
where $\beta \in [0,1]$ is the leak factor. A spike is emitted whenever the membrane potential exceeds a threshold $\theta$:
\begin{equation}
o[t] =
\begin{cases}
1, & u[t] \geq \theta, \\
0, & \text{otherwise},
\end{cases}
\end{equation}
after which the membrane is reset by subtracting $\theta$ from $u[t]$. Since the spiking function is non-differentiable, the network is trained via surrogate gradient learning \cite{fang2021deep}, implemented using the snnTorch package \cite{eshraghian2021training}.

\subsection{Spike Decoding}

The spikes $o[t]$ from the network's final layer are averaged over the $N$ timesteps and mapped to an unbounded estimate via the logit function,
\begin{equation}
\bar{o} = \mathrm{logit}\left(\frac{1}{N}\sum_{t=1}^{N}o[t]\right).
\end{equation}
The averaged output is clipped to $[\epsilon,1-\epsilon]$ before the logit. This is applied to each of the network's two output channels to form the complex mask estimate $M[k,\ell]$ of Section~3.
\tikzset{
tensor/.style={draw, thick, rounded corners=1pt, minimum width=2.9cm, minimum height=0.70cm, align=center, fill=yellow!55},
process/.style={draw, thick, rounded corners=1pt, minimum width=2.9cm, minimum height=0.70cm, align=center, fill=teal!55},
activation/.style={draw, thick, rounded corners=1pt, minimum width=2.0cm, minimum height=0.50cm, align=center, fill=orange!45},
snnblock/.style={draw, thick, rounded corners=1pt, minimum width=2.9cm, minimum height=0.70cm, align=center, fill=red!35}
}

\begin{figure}[t]
\centering
\resizebox{\linewidth}{!}{
\begin{tikzpicture}[font=\scriptsize, node distance=0.20cm, >=latex]
 
\node[draw, dashed, thick, rounded corners=10pt, inner sep=0.25cm] (annbox)
{
\begin{tikzpicture}[font=\scriptsize, node distance=0.20cm, >=latex]
\node[tensor] (input) {Input STFT Features\\$[T,F,2C{+}2]$};
\node[process, below=of input] (fc1) {Linear Layer\\$[T,F,H]$};
\node[activation, below=of fc1] (relu1) {ReLU};
\node[process, below=of relu1] (fc2) {Linear Layer\\$[T,F,H/2]$};
\node[activation, below=of fc2] (relu2) {ReLU};
\node[process, below=of relu2] (linear) {Output Linear Layer\\$[T,F,2]$};
\node[tensor, below=of linear] (output) {Complex Mask\\$[T,F]$};
\draw[->, thick] (input) -- (fc1);
\draw[->, thick] (fc1) -- (relu1);
\draw[->, thick] (relu1) -- (fc2);
\draw[->, thick] (fc2) -- (relu2);
\draw[->, thick] (relu2) -- (linear);
\draw[->, thick] (linear) -- (output);
\end{tikzpicture}
};
\node[font=\bfseries\small, above=0.15cm of annbox] {FC-ANN};
 
\node[draw, dashed, thick, rounded corners=10pt, inner sep=0.25cm, right=1.2cm of annbox] (snnbox)
{
\begin{tikzpicture}[font=\scriptsize, node distance=0.20cm, >=latex]
\node[tensor] (input) {Input STFT Features\\$[T,F,2C{+}2]$};
\node[snnblock, below=of input] (encode) {Spike Encoding\\$[N,T,F,2C{+}2]$};
\node[process, below=of encode] (fc1) {Linear Layer\\$[N,T,F,H]$};
\node[activation, below=of fc1] (lif1) {LIF Neurons};
\node[process, below=of lif1] (fc2) {Linear Layer\\$[N,T,F,H/2]$};
\node[activation, below=of fc2] (lif2) {LIF Neurons};
\node[process, below=of lif2] (linear) {Output Linear Layer\\$[N,T,F,2]$};
\node[activation, below=of linear] (lif3) {LIF Neurons};
\node[snnblock, below=of lif3] (decode) {Spike Decoding\\$[T,F,2]$};
\node[tensor, below=of decode] (output) {Complex Mask\\$[T,F]$};
\draw[->, thick] (input) -- (encode);
\draw[->, thick] (encode) -- (fc1);
\draw[->, thick] (fc1) -- (lif1);
\draw[->, thick] (lif1) -- (fc2);
\draw[->, thick] (fc2) -- (lif2);
\draw[->, thick] (lif2) -- (linear);
\draw[->, thick] (linear) -- (lif3);
\draw[->, thick] (lif3) -- (decode);
\draw[->, thick] (decode) -- (output);
\end{tikzpicture}
};
\node[font=\bfseries\small, above=0.15cm of snnbox] {FC-SNN};
\end{tikzpicture}
}
\caption{FC-ANN baseline architecture and its equivalent SNN implementation (FC-SNN). The SNN architecture replaces ReLU activations with leaky integrate-and-fire (LIF) neurons and includes spike encoding and decoding stages. $C$ denotes the number of microphones, $H$ and $H/2$ denote hidden feature dimensions, and $N$ denotes the number of spike timesteps.}
\label{fig:ann_snn_architectures}
\end{figure}
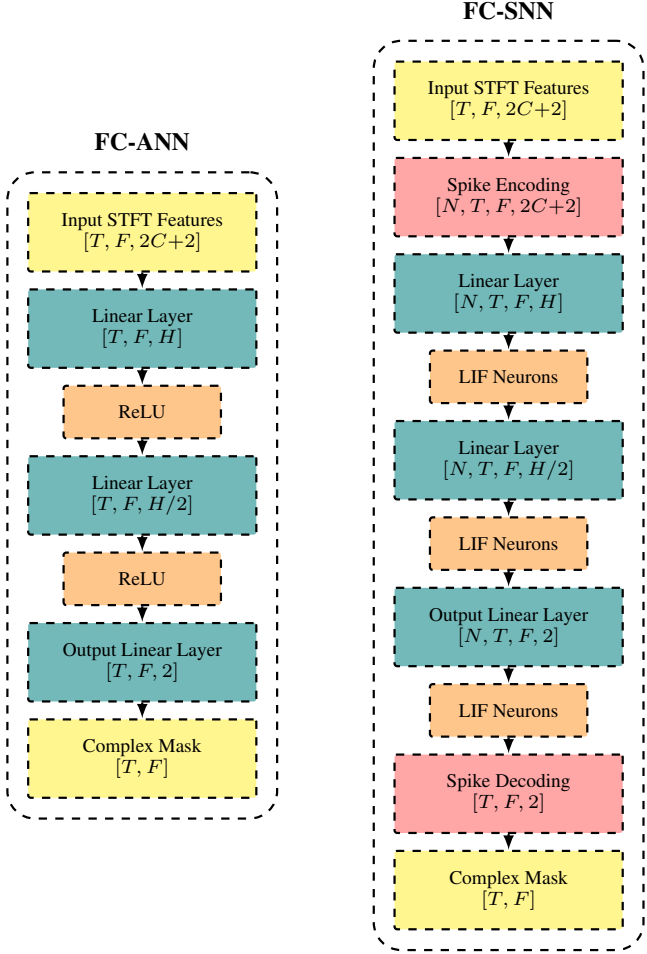

\section{Experimental Evaluation}
This section evaluates FC-SNN's performance and compares it against the baselines of Section~3.
\subsection{Experimental Setup}

\textbf{Dataset.} Following the simulation methodology of \cite{tesch2022insights}, we simulate a semi-circular 6-microphone array (8~cm radius) spanning microphone angles $-180^\circ$ to $0^\circ$, recording one target speaker fixed at a $0^\circ$ azimuth relative to the array (arbitrary target directions could be supported via target-direction conditioning \cite{tesch2023multi}), and five interfering speakers positioned around the array, with room impulse responses generated via the image-source method (\texttt{gpuRIR}~\cite{diaz2021gpurir}). The dataset comprises 6000 training, 1000 validation, and 600 test mixtures (up to 600 per epoch for training/validation), each cropped to a fixed 6-second segment starting from the onset of the target speech. Additional parameters are given in Table~\ref{tab:simulation_parameters}.

\noindent\textbf{Training.} All models are trained with Adam (batch size 1, learning rate $10^{-3}$) for 250 epochs to minimize $\mathcal{L} = -\mathrm{SI\text{-}SDR}(\hat{s},s)$; the checkpoint with the lowest validation loss is retained for evaluation.

\noindent\textbf{Metrics.} We report SI-SDR, PESQ, and STOI, together with parameter count and computational complexity: multiply-accumulate operations (MACs) for densely-computed ANN and non-spiking layers, and spike-driven synaptic operations (SynOps) for genuinely spike-driven layers. MACs and SynOps reflect different underlying hardware operations and are not directly comparable as an energy metric without a target hardware implementation; they are reported separately as complexity indicators rather than converted to a common energy unit. Under direct encoding, the first layer computes an identical output at every one of the $N$ timesteps, since its input does not vary; we therefore additionally report an \textit{effective} MAC rate, obtained by dividing the as-implemented MAC rate by $N$, reflecting the cost of an implementation that computes this layer once and reuses the result rather than recomputing it $N$ times.

\begin{table}[t]
\centering
\caption{Experimental parameters.}
\label{tab:simulation_parameters}
\begin{tabular}{ll}
\hline
Parameter & Value \\
\hline
Speech corpus & WSJ0 \\
Sampling frequency & 16 kHz \\
Interfering speakers & 5 \\
Array geometry & Semi-circular \\
Number of microphones & 6 \\
Array radius & 8 cm \\
Room dimensions & $2.5$--$5.0 \times 3.0$--$9.0 \times 2.2$--$3.5$ m \\
Reverberation time ($T_{60}$) & $0.2$--$0.5$ s \\
STFT & 512 FFT, 32 ms window, 16 ms hop \\

\hline
\end{tabular}
\end{table}

\subsection{Model Size Scaling}

Table~\ref{tab:model_size_scaling} reports SI-SDR, PESQ, and STOI as a function of hidden width $H$ and encoding scheme for FC-SNN (direct: $N=20$; rate: $N=100$), alongside the FC-ANN/FT-JNF baselines. FT-JNF peaks at $H=128$, with lower performance on either side, motivating its use as the compact reference point in Table~\ref{tab:main_results}. Direct encoding does not converge at $H=512$, while rate encoding converges across all model sizes. At $H=128$ rate encoding falls behind direct encoding by 2.07~dB in SI-SDR, 0.124 in PESQ, and 0.036 in STOI. Overall, performance improves only moderately with increasing model size under both encoding schemes.

\begin{table}[t]
\centering
\resizebox{\linewidth}{!}{
\begin{tabular}{llccccc}
\hline
\textbf{Architecture} & \textbf{Enc.} & \textbf{$H$} & \textbf{Params} & \textbf{SI-SDR (dB)} & \textbf{PESQ} & \textbf{STOI} \\
\hline
FC-SNN & direct & 64  & 3.1K  & 2.79 & 1.538 & 0.797 \\
FC-SNN & direct & 128 & 10.3K & 3.14 & 1.550 & 0.805 \\
FC-SNN & direct & 256 & 37.0K & 3.42 & 1.607 & 0.813 \\
FC-SNN & direct & 512 & --     & \multicolumn{3}{c}{convergence failure} \\
FC-SNN & rate   & 64  & 3.1K   & 0.78 & 1.408 & 0.763 \\
FC-SNN & rate   & 128 & 10.3K  & 1.07 & 1.426 & 0.769 \\
FC-SNN & rate   & 256 & 37.0K  & 1.21 & 1.425 & 0.774 \\
FC-SNN & rate   & 512 & 139.5K & 1.28 & 1.423 & 0.775 \\
\hline
FC-ANN & --     & 64  & 3.1K    & 3.95 & 1.557 & 0.819 \\
FC-ANN & --     & 128 & 10.3K   & 4.80 & 1.626 & 0.836 \\
FC-ANN & --     & 256 & 37.0K   & 5.07 & 1.657 & 0.843 \\
FC-ANN & --     & 512 & 139.5K  & 5.39 & 1.678 & 0.848 \\
\hline
FT-JNF & --     & 64  & 81.5K     & 6.67 & 1.843 & 0.879 \\
FT-JNF & --     & 128 & 310.5K   & 7.83 & 2.019 & 0.900 \\
FT-JNF & --     & 256 & 1.21M & 7.59 & 1.995 & 0.896 \\
\hline
\end{tabular}
}
\caption{Enhancement performance as a function of model size (hidden width $H$) and encoding scheme (direct: fixed $N=20$; rate: fixed $N=100$).}
\label{tab:model_size_scaling}
\end{table}

\begin{table*}[t]
\centering
\begin{tabular}{clcccccc}
\hline
\textbf{$N$} & \textbf{Enc.} & \textbf{SI-SDR (dB)} & \textbf{PESQ} & \textbf{STOI} & \textbf{GMAC/s} & \textbf{Eff. GMAC/s} & \textbf{GSynOp/s} \\
\hline
10  & direct & 2.89  & 1.544 & 0.800 & 0.577 & 0.058 & 0.685 \\
20  & direct & 3.42 & 1.607 & 0.813 & 1.154 & 0.058 & 1.311 \\
50  & direct & 4.03  & 1.629 & 0.826 & 2.886 & 0.058 & 3.373 \\
10  & rate   & -1.92 & 1.259 & 0.693 & -- & -- & 0.87 \\
100 & rate   & 1.21  & 1.425 & 0.774 & -- & -- & 10.18 \\
200 & rate   & 1.88  & 1.459 & 0.784 & -- & -- & 22.61 \\
\hline
\end{tabular}
\caption{Enhancement performance and computational complexity as a function of spike-train length $N$ and encoding scheme, for fixed model size $H=256$.}
\label{tab:spike_length_scaling}
\end{table*}
\begin{table*}[t]
\centering
\begin{tabular}{lccccccc}
\hline
\textbf{Method} & \textbf{SI-SDR (dB)} & \textbf{PESQ} & \textbf{STOI} & \textbf{Params} & \textbf{GMAC/s} & \textbf{Eff. GMAC/s} & \textbf{GSynOp/s} \\
\hline
Unprocessed     & -7.18 & 1.14  & 0.576 & --     & --   & --    & -- \\
MVDR (oracle)   & 2.06  & 1.326 & 0.827 & --     & --   & --    & -- \\
FT-JNF          & 7.83  & 2.019 & 0.900 & 310.5K & 4.95 & --    & -- \\
FC-ANN          & 5.07  & 1.657 & 0.843 & 37.0K  & 0.59 & --    & -- \\
\hline
FC-SNN (direct) & 4.03  & 1.629 & 0.826 & 37.0K  & 2.89 & 0.058 & 3.37 \\
FC-SNN (rate)   & 1.21  & 1.425 & 0.774 & 37.0K  & --   & --    & 10.18 \\
\hline
\end{tabular}
\caption{Performance and computational complexity comparison of the proposed SNN architecture against baseline methods.}
\label{tab:main_results}
\end{table*}

\subsection{Spike-Train Length Scaling}

Fixing $H=256$, Table~\ref{tab:spike_length_scaling} reports performance and computational complexity as a function of spike-train length $N$ for both encoding schemes. Under direct encoding, the first layer is computed densely (non-zero GMAC/s); its effective MAC rate (Section~5.1) is constant at 0.058~GMAC/s regardless of $N$, since the as-implemented cost scales linearly with $N$ while the underlying computation does not. Under rate encoding every layer is genuinely spike-driven, so no MAC operations occur at all (GMAC/s and effective GMAC/s are both not applicable). Spike-train length has a substantial effect on enhancement quality under both schemes: at $N=10$, direct encoding outperforms rate encoding substantially, and rate encoding's SI-SDR increases monotonically with $N$ across the full range tested (10 through 200), requiring roughly an order of magnitude more timesteps ($N=200$: 1.88~dB) yet still not matching direct encoding's $N=10$ result (2.89~dB).

\subsection{Operating Point Selection}

Combining the model-size and spike-train-length results across both encoding schemes, we select $H=256$ as the representative model size for both. Direct encoding at $N=50$ provides the best enhancement quality among the configurations considered (4.03~dB SI-SDR), while rate encoding at $N=100$ provides full spike-driven computation at moderate complexity. These two operating points, carried forward to the baseline comparison in Table~\ref{tab:main_results}, illustrate the core trade-off of this work: direct encoding when raw performance is the priority, rate encoding when full spike-driven computation is required.

\subsection{Comparison with Baseline Methods}

Table~\ref{tab:main_results} compares the selected direct- and rate-encoded FC-SNN operating points against the unprocessed mixture, oracle MVDR, FT-JNF, and FC-ANN. As-implemented, direct-encoding FC-SNN's 2.89~GMAC/s exceeds FC-ANN's 0.59~GMAC/s; under the effective-MAC accounting (Section~5.1), however, its MAC rate drops to 0.058 GMAC/s, roughly 10 times lower than FC-ANN. While both SNN variants substantially reduce parameter count relative to FT-JNF, recent SNN-based speech enhancement models are substantially larger: Spiking-S4~\cite{du2024spiking} uses 530K parameters, Spiking-FullSubNet~\cite{hao2025toward} 965K, and DPSNN~\cite{sun2024dpsnn} up to 1.4M, all at least an order of magnitude larger than FC-SNN's 37.0K ($H=256$). Overall, these results highlight FC-SNN as a compact approach for spike-based multichannel speech enhancement.

\section{Conclusion}

We presented a fully spiking framework for multichannel speech enhancement based on a fully connected SNN architecture, derived from a compact ANN baseline via leaky integrate-and-fire neurons. We characterized how enhancement performance scales with model size and spike-train length relative to ANN and classical beamforming baselines, reporting parameter counts and computational complexity. Comparing direct and rate spike encoding revealed a clear trade-off: direct encoding achieves substantially higher enhancement quality per timestep, while rate encoding trades this for full spike-driven computation across every layer of the network.

\footnotesize
\bibliographystyle{IEEEbib}
\bibliography{bib1}
\end{document}